\documentclass[10pt,showpacs,twocolumn,superscriptaddress,aps,prb]{revtex4-1}
\usepackage{amsmath,amssymb,graphicx}
\pdfoutput=1 

\begin{document}
\title{Modelling and experiments of self-reflectivity under femtosecond ablation conditions}
\author{H. Zhang}\email{Electronic mail: zhterran@qq.com}
\affiliation{Debye Institute for Nanomaterials Science, Utrecht University, P. O. Box 80000, 3508 TA Utrecht, The Netherlands}
\author{S. A. Wolbers}
\affiliation{Debye Institute for Nanomaterials Science, Utrecht University, P. O. Box 80000, 3508 TA Utrecht, The Netherlands}
\author{D. M. Krol}
\affiliation{Debye Institute for Nanomaterials Science, Utrecht University, P. O. Box 80000, 3508 TA Utrecht, The Netherlands}
\affiliation{University of California Davis, Davis, CA 95616}
\author{J. I. Dijkhuis}
\affiliation{Debye Institute for Nanomaterials Science, Utrecht University, P. O. Box 80000, 3508 TA Utrecht, The Netherlands}
\author{D. van Oosten}\email{Corresponding author: D.vanOosten@uu.nl}
\affiliation{Debye Institute for Nanomaterials Science, Utrecht University, P. O. Box 80000, 3508 TA Utrecht, The Netherlands}

\begin{abstract}
We present a numerical model which describes the propagation of a single femtosecond laser pulse in a medium of which the optical properties dynamically change within the duration of the pulse. We use a Finite Difference Time Domain (FDTD) method to solve the Maxwell's equations coupled to equations describing the changes in the material properties. We use the model to simulate the self-reflectivity of strongly focused femtosecond laser pulses on silicon and gold under laser ablation condition. We compare the simulations to experimental results and
find excellent agreement. 
\end{abstract}
\maketitle

\section{Introduction}
\label{Sec:Intro}
Recent advances in ultrafast laser material processing enable nano-sized structures to be directly fabricated in various materials~\cite{Gamaly2006,Liao2013,HaoAPL2011,Li2003}. The basic processes during femtosecond laser ablation are absorption by electrons, energy transfer to the lattice and subsequent
material removal. These processes are all temporally well separated~\cite{Rethfeldtimescales2004}. 
The first step, the absorption of light, is a complex and interesting problem.
As the laser pulses are focused into a spot size comparable to the wavelength of the laser light,
the interaction between the light and the material takes place in a very confined volume. 
Furthermore, due to the high peak intensities involved, nonlinear optical effects play a dominant role. 
Typically, when a laser pulse propagates through a semiconductor or an insulator, multi-photon absorption 
takes place during the leading part of the pulse. This leads to the generation of a high concentration of
charge carriers. When a laser pulse impinges on a metal, the existing free electrons in the metal will be 
strongly heated by the leading part of the pulse. In both cases the leading part of the pulse alters the optical properties of the material, which implies that the trailing part of the laser pulse interacts with a material whose optical properties are significantly different from those of the unexcited material. A detailed numerical modeling of this process provides insights into the complex mechanism of energy deposition under these conditions and is therefore crucial to describe laser nano-processing.

In earlier one-dimensional models of the absorption of femtosecond laser pulses in silicon, the dynamically 
changing optical properties were taken into account using a nonlinear Lambert-Beer law,
with absorption coefficients that change dynamically due to the generation of free carriers. The results of 
these models are in good agreement with reflectivity measurements performed with weakly focused laser beams~
\cite{Hulin1984,Tinten2000,Tinten1995}. However, these one-dimensional models are expected not to be 
adequate to describe the laser-matter interaction in sub-wavelength volumes and with the large focusing 
angles obtained with high numerical aperture objectives.
In other studies, the propagation of femtosecond laser pulses in nonlinear media is simulation by solving 
the non-linear Schr\"odinger equation (NLSE)~\cite{Boyd2007,Brabec1997}. 
However, the self-scattering by the sub-wavelength plasma formed during nano-ablation~\cite{Zhang2013} implies the breakdown of the slowly-varying-amplitude-approximation on which the NLSE is based. 
Additionally, the high numerical aperture objective used to focus the beam implies the breakdown of the paraxial approximation, another approximation used in the derivation of the NLSE. 
Due to these limitations of the NLSE, there is an increasing interest in the development of numerical models 
that resolve the full set of Maxwell's equations coupled to the equations describing the changes in the materials induced by the pulse under tight focusing conditions~\cite{Ludovic2007,Mezel2008,Bogatyrev2011,Schmitz2012}. However existing studies do not address laser-matter interaction in metals, do not find quantitative agreement with experiments or lack a comparison to experiments. 

In this paper, we present a numerical model of the laser energy deposition in femtosecond laser nano-
processing of both semiconductors and metals. The model simulates the propagation of light using a two-dimensional Finite Difference Time Domain (FDTD) method, coupled to a set of differential equations that 
describe the changes in the material properties that are driven by the laser light. The model is 
compared with self-reflectivity measurements of a strongly focused femtosecond laser beam in single-shot 
ablation experiments on silicon and gold. We show that the model excellently describes the self-reflectivity measurements on the four types of specimens we investigated, namely two silicon-on-insulator samples with different device layer thickness, bulk silicon and gold. We further show that in the case of strong 
focusing, a one-dimensional model does not reproduce the experimental results and that a two-dimensional model is thus required. As our model excellently agrees with the experimental results, without the use of fitting parameters, it can be used to study and optimize the energy deposition in femtosecond 
laser nano-processing of materials.

The paper is organized as follows. In Section~\ref{Sec:theory} we discuss the theoretical model we use to describe the propagation of an intense laser pulse in a medium of which the optical properties are changing during the pulse. In Section~\ref{Sec:model}, we describe in detail the numerical implementation of that model. In Section~\ref{Sec:Exp_and_model}, we compare the results of the model to experimental results.
A summary and conclusion are presented in Section~\ref{Sec:Summary}.

\section{Theory}
\label{Sec:theory}
The propagation of electromagnetic waves is in general governed by Maxwell's equations
\begin{equation*}
\nabla\times \mathbf{H}=\frac{\partial \mathbf{D}}{\partial t}+\mathbf{j},\,\,
\nabla\times \mathbf{E}=-\frac{\partial \mathbf{B}}{\partial t},
\label{eqn:Maxwell's equations 1}
\end{equation*}
in combination with relations defining the auxiliary fields
\begin{equation*}
\mathbf{D}=\epsilon_0\mathbf{E}+\mathbf{P},\,\,
\mathbf{H}=\frac{1}{\mu_0}\epsilon_0\mathbf{B}-\mathbf{M}.
\label{eqn:auxiliary_Maxwell's equations}
\end{equation*}
The above relations are specific to the medium and can be modelled using microscopic theories. In a linear and homogeneous medium, we can write $\mathbf{P}=\epsilon_0\chi\mathbf{E}$ and \linebreak $\mathbf{M}=\chi_m\mathbf{B}/\mu_0$, where $\chi$ is the electric susceptibility and $\chi_m$ is the magnetic susceptibility. For electromagnetic waves at optical frequencies, the magnetic susceptibility is almost always negligibly small. We will therefore neglect it in the remainder of this paper and only consider the electric susceptibility. In the simple linear and homogeneous case, the above set of equations can be easily cast into a wave equation which can then be solved analytically. However, in relevant cases, the susceptibility is not homogeneous. In that situation, one in general needs numerical methods to solve Maxwell's equations. Furthermore, when sufficiently strong fields are applied, nonlinear effects can come into play. For example, absorption of light by the medium can result in the generation of free charge-carriers. These free carriers will contribute to the susceptibility. As this will make the susceptibility depend on the intensity of light inside the medium, free-carrier generation also leads to inhomogeneity even in case of an initially homogeneous medium. Local heating of the material has a similar effect, as it locally changes the refractive index and thus makes the medium inhomogeneous. 

\begin{figure}[h]
\begin{center}
\includegraphics[width=7cm]{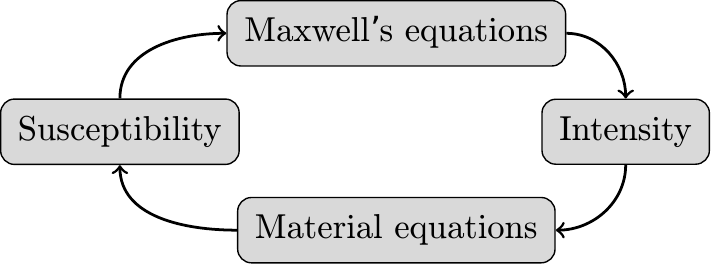}
\end{center}
\caption{Diagramatic view of the model}
\label{fig:simple_model}
\end{figure}
In many cases, the change in the local susceptibility are small and slow enough to be negligible when considering the propagation of light. However, when using focused pico- or femtosecond pulses, the changes in local susceptibility occur during the pulse itself. As illustrated in Fig.~\ref{fig:simple_model}, this requires one to solve the equations governing the dynamics of the properties of the material, which is driven by the intensity of light. This intensity 
is obtained by solving the Maxwell's equations, which use the material properties as an input.

If we assume that the susceptibility changes slowly with respect to the oscillation period of the light,
we can neglect time derivatives of the susceptibility and thus write the time derivative of the 
displacement field as 
$\partial {\bf D}/\partial t=(1+\chi) \partial {\bf E}/\partial t$.
This reduces the Maxwell's equations to
\begin{eqnarray}
\label{eqn:maxwell_1}
\frac{\partial {\bf H}({\bf r},t)}{\partial t}&=&\frac{1}{\mu_0}\nabla\times{\bf E}({\bf r},t),\\
\label{eqn:maxwell_2}
\frac{\partial{\bf E}({\bf r},t)}{\partial t}&=&\frac{1}{\varepsilon_0\varepsilon_r({\bf r},t)}\nabla\times{\bf H}({\bf r},t),
\end{eqnarray}
where we have introduced the dielectric function $\varepsilon_r({\bf r},t)=1+\chi({\bf r},t)$. 

To determine the susceptibility, we need to determine certain position- and time-dependent properties 
of the material. For instance, we require the carrier density $N({\bf r},t)$, which we can obtain by 
integrating the diffusion equation
\begin{equation}
\frac{\partial N({\bf r},t)}{\partial t}+\nabla\cdot\left[-D_0({\bf r},t)\nabla N({\bf r},t)\right]=S_N({\bf E}({\bf r},t)),
\label{eqn:carrier_density_1}
\end{equation}
where the source term $S_N$ depends on the intensity of light (and thus on the electric-field amplitude) as 
absorption of light leads to the formation of free carriers. 

The carrier density distribution obtained from the diffusion equation is subsequenty used to calculate 
position- and time-dependent susceptibility (and thus the dielectric function)
\begin{equation}
\varepsilon_r({\bf r},t)=1+\chi({\bf r},N({\bf r},t),...),
\end{equation}
where we have now explicitly written the susceptibility as a function of the carrier density. The dots
indicate that the susceptibility is also a function of other properties of the medium, such as electron temperature, lattice temperature, etc.. If these are expected to vary on the timescale of the pulse, 
additional equations need to be include to describe their dynamics. 

In the next section, we will follow the outline given above to derive a model to describe the 
self-reflectivity of a semiconductor structure subjected to intense femtosecond laser illumination. 
We will explicitely discuss the physical processes that should be taken into account and give details of
the numerical implementation of the model.

\section{The model}
\label{Sec:model}

As described in the previous section, we need to solve a set of equations to describe how the 
dielectric function evolves during the laser pulse. Which equations we need to solve depends on the
material used. Here we describe the relevant sets of equations required to model the self-reflectivity 
of a semiconductor as well as a metal under ablation conditions and show that the model is in excellent agreement with experimental results for both types of materials.

\subsection{Laser-matter interaction for silicon}
\label{subsec:LASER-MATTER INTERACTION FOR SILICON}

\begin{figure}[tbp]
   \centering
   \includegraphics[width=8.4cm]{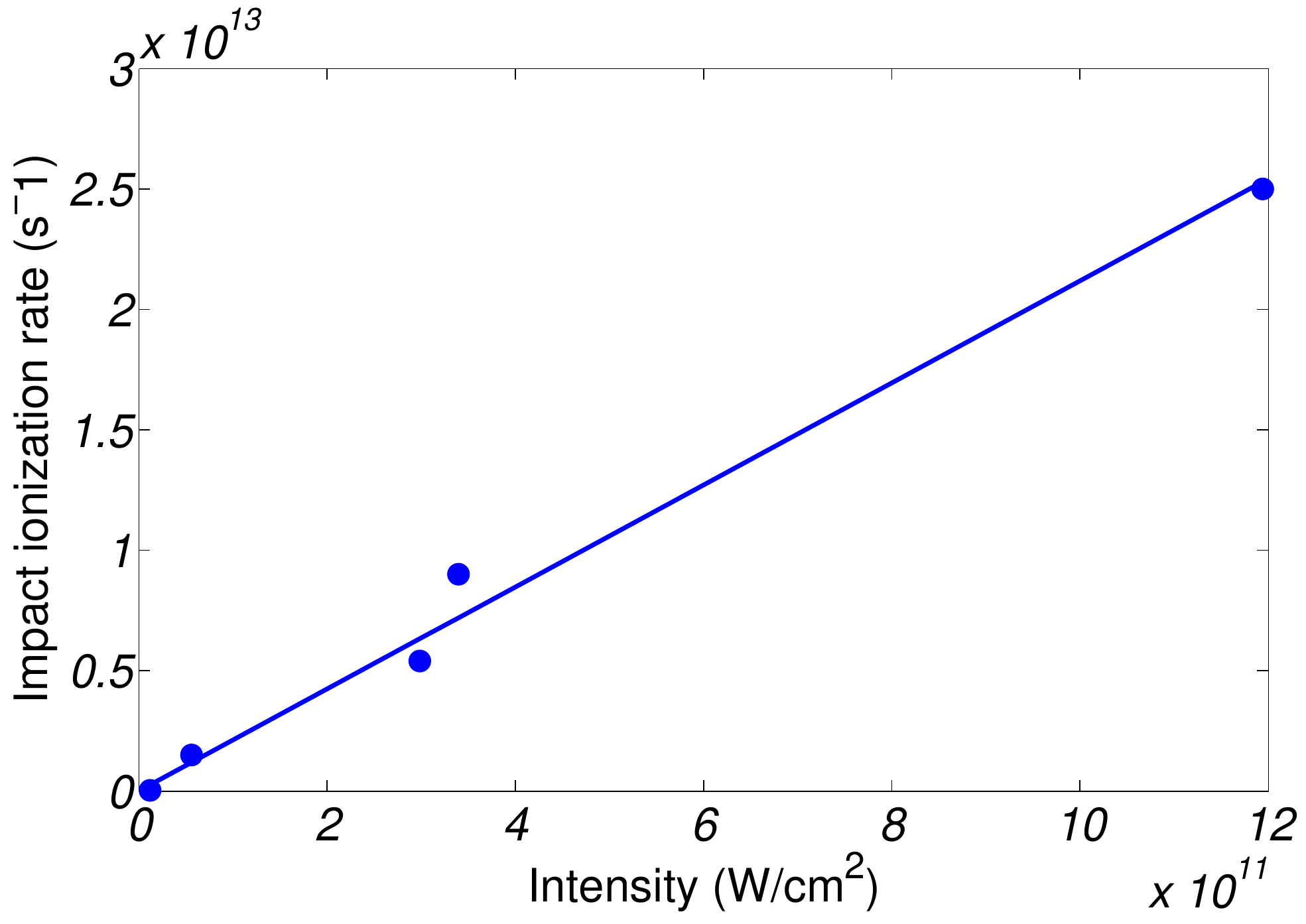}
   \caption{Impact ionization rate in silicon. An impact ionization rate $\rm \theta=21.2~cm^2/J$ is obtained by a linear fit of the data presented in Ref.~\cite{Pronko1998}.}
\label{fig:impact_rate_silicon}
\end{figure}

The first process during the absorption of a femtosecond laser pulse by a semiconductor is the excitation of electrons from the valence band to the conduction band. Depending on the band gap of the material and the incident photon energy, the excitation can be either one-photon absorption or multi-photon absorbtion or both. Subsequently, electrons already excited to the conduction band can gain energy in the laser field via free-carrier absorption. If the excess energy of a conduction electron is sufficiently high, it may excite another electron in the valence band to the conduction band, by a process known as impact ionization. When this process occurs multiple times, it is referred to as avalanche or cascade ionization. 
For silicon at $800~\rm nm$ excitation wavelength, free carriers are created by one-photon absorption (OPA), two-photon absorption (TPA)~\cite{Choi2002,Tinten2000}, and impact ionization~\cite{Pronko1998}. 
Generally, impact ionization is a very complicated process that involves the energy distribution of the electrons~\cite{Medvedev2010}. Here we use a simplified and convenient expression for the impact ionization term, deduced by Stuart {\it et.~al.}~\cite{Stuart1996}. We rewrite Eq.~(\ref{eqn:carrier_density_1}) as 
\begin{eqnarray}
\frac{\partial N({\bf r},t)}{\partial t}+\nabla\cdot[-D\nabla N({\bf r},t)]&=&\frac{\alpha_0 I({\bf r},t)}{\hbar\omega}+\frac{\beta I^2({\bf r},t)}{2\hbar\omega}\nonumber\\
&+&\theta I({\bf r},t)N({\bf r},t),
\label{eqn:carrier_gen}
\end{eqnarray}
where $D$ is the carrier diffusivity, $\alpha_0$ is the OPA coefficient, $\beta$ is the TPA coefficient, and $\theta$ is the impact ionization coefficient.
The intensity $I({\bf r},t)$ that appears in the source term on the right-hand side of the above equation is the laser intensity inside the medium, which can be obtained from the amplitude of the time-varying electric field in the material using
\begin{equation}
I({\bf r},t)=\frac{1}{2}\epsilon_0 c\rm {Re}\rm\{n({\bf r},t)\}\left| E_0({\bf r},t)\right|^2,
\label{eqn:I_gen}
\end{equation}
where ${\rm Re}\{n\}$ denotes the real part of the refractive index $n$. 
We deduce the impact ionization coefficient $\theta$ from the experimental results obtained by Pronko~{\it et al.} who measured the impact ionization rate ($s^{-1}$) for the dielectric breakdown in silicon with $786~\rm nm$ fs laser pulses.~\cite{Pronko1998} In Fig.~\ref{fig:impact_rate_silicon}, we plot their results as a function of intensity. Fitting the data yields an impact ionization coefficient $\theta=21.2~\rm cm^2/J$.
Due to the creation of a high density of carriers, the optical response of silicon under ablation conditions 
is dominated by the free-carrier response, which can be calculated using the Drude model~\cite{Tinten2000,Choi2002}. In our model, we also take into account the changes in the dielectric constant due to 
the optical Kerr effect and TPA. Thus the dielectric function of strongly excited silicon 
($\epsilon_{\rm ex}$) can be written as
\begin{eqnarray}
\epsilon_{\rm ex}({\bf r},t)&=&\epsilon_{\rm Si}+\epsilon_{\rm Drude}+\epsilon_{\rm NL},\\
\epsilon_{\rm Drude}({\bf r},t)&=&-{\frac{\left(\omega_p/\omega\right)^2}{1+i/\omega\tau_d}},\\
\epsilon_{\rm NL}({\bf r},t)&=&\frac{3}{4}\chi_3{\lvert E_0(r,t)\rvert}^2,\\
\omega_p({\bf r},t)&=&\sqrt{\frac{N({\bf r},t)e^2}{m^*m_e\epsilon_0}},
\label{eqn:epsilon_excited_silicon}
\end{eqnarray}
where $\epsilon_{\rm Si}$ is the dielectric constant of unexcited silicon at $\rm 800~nm~(13.6+0.048i)$~\cite{Choi2002} and $\tau_d$ is the carrier collision time, which is believed to be around $\rm 1~fs$~\cite{Hulin1984,Tinten2000} at the excitation level relevant for this work. Finally $m^*m_e$ is the optical effective mass of highly excited silicon, which we will discuss later in this section. The third-order susceptibility $\chi_3$ can be calculated from the value of the Kerr coefficient $n_2$ and the value of TPA 
coefficient $\beta$ in silicon. The refractive index $n$ and the effective absorption coefficient $\alpha_{\rm ex}$ of the excited material are
\begin{eqnarray}
n_{\rm ex}&=&\sqrt{\epsilon_{\rm ex}},\\
\alpha_{\rm ex}&=&\frac{4\pi {\rm Im}\rm\{n_{\rm ex}\}}{\lambda_0}.
\label{eqn:n_excited_silicon}
\end{eqnarray}
Although the carrier temperature does not explicitely appear in the above equations for the dielectric
function, we nevertheless have to calculate it. This is because the carrier diffusivity $D$ and the
optical effective mass $m^* m_e$ depend on the carrier temperature. To model the change of the carrier temperature $T_c$ during the pulse, we use a heat equation~\cite{Choi2002,Tinten2000}
\begin{equation}
\frac{\partial [U_c({\bf r},t)]}{\partial t}+\nabla\cdot[-\kappa_c\nabla
T_c({\bf r},t)]=\alpha_{\rm ex} I({\bf r},t).
\label{eqn:electron_temperature}
\end{equation}
The total energy density $U_c$ of the excited carriers is given by
\begin{equation}
Uc=C_cT_c+NE_g,
\label{eqn:carrier_total_energy}
\end{equation}
with $C_c$ the carrier heat capacity, $E_g$ the band gap energy and $\kappa_c$ the thermal conductivity 
of the carriers. As the carrier temperature reached at the excitation level relevant to this work exceeds $\rm 10^4~K$, the induced plasma is nondegenerate even at high densities~\cite{Tinten2000}. Thus, the carrier heat capacity and the carrier heat conductivity can be approximated using classical thermodynamics~\cite{Ashcroft1976},
\begin{eqnarray}
C_c&=&3k_BN({\bf r},t),\\
\kappa_c&=&\frac{1}{3}C_c\langle v_c\rangle^2\tau_d,
\label{eqn:carrier heat conductivity}
\end{eqnarray}
where $k_B$ is the Boltzmann constant, $\langle v_c\rangle=\sqrt{3k_BT_c/m^*}$ the thermal velocity of carriers, $m^*$ the carrier effective mass and $\tau_d$ the carrier-carrier collision time. The carrier diffusivity is then found using the Einstein relation~\cite{Ashcroft1976}
\begin{equation}
D=\frac{k_BT_c\tau_d}{m^*}.
\label{eqn:Einstein_relation}
\end{equation}
In Ref.~\cite{Tinten2000} the authors determine a static optical effective mass of the femtosecond laser-induced plasma in silicon using time-resolved pump-probe experiments. However, as the optical effective mass is both temperature and density dependent, it will actually change during the pulse. For the carrier temperatures relevant for this work, the Fermi-Dirac reduces to a Boltzmann distribution, resulting in an effective mass that is independent of the carrier density~\cite{Riffe2002}. The temperature dependence remains and is easy to understand; when the electrons are heated up far beyond the conduction band edge, the curvature of the band decreases, giving rise to a larger effective mass.  In this case, the optical effective mass increases approximately linearly with carrier temperature~\cite{Riffe2002}. As 
experimental measurements of the optical effective mass are based on the Drude model, they only depend on 
the ratio $N/m^*$ (see Eq.~(\ref{eqn:epsilon_excited_silicon})). So to measure $m^*$ one needs to know the carrier density $N$, which in our case is a-priori unknown. Therefore, we use the relation suggested by Riffe's theoretical calculation which takes the detailed band structure of silicon into account. His calculation shows that the optical effective mass can in our regime be approximated by
\begin{equation}
m^*=m^*_0+m_kT_c,
\label{eqn:temperture_dependent_effective_mass}
\end{equation}
where the optical effective mass of unperturbed silicon $m^*_0$ is well-known to be $0.15~m_e$~\cite{Tinten2000,Riffe2002}. We extract the slope $m_k=3.1\times10^{-5}~\rm K^{-1}$ from Riffe's calculations~\cite{Riffe2002}. 

\subsection{Laser-matter interaction for gold}
\label{subsec:LASER-MATTER INTERACTION FOR GOLD}

In the case of femtosecond laser ablation of gold, the optical absorption process is somewhat different. 
As there is already a high density of free electrons present in the unexcited material, the number of free 
electrons is not significantly influenced by the laser pulse. 
The dominant absorption mechanism is therefore the heating of those electrons. In analogy with Eq.~(\ref{eqn:electron_temperature}) we use
\begin{equation}
C_e\frac{\partial [T_e({\bf r},t)]}{\partial t}+\nabla\cdot[-\kappa_e\nabla
T_e({\bf r},t)]=\alpha_{\rm ex} I({\bf r},t),
\label{eqn:electron_temperature_gold}
\end{equation}
where the subscript $e$ denotes that the only charge carriers are electrons. At the high electron 
temperatures reached during ablation, we can approximate the electronic heat capacity as
\begin{equation}
C_e=\frac{3}{2}k_BN,
\label{eqn:electron heat capacity}
\end{equation}
where the electron density $N$ is now kept constant during the simulation.
The change of electron temperature gives rise to a change in the Drude damping time which is given by
\begin{equation}
\frac{1}{\tau_d}=\frac{1}{\tau_{e-e}}+\frac{1}{\tau_{e-l}},
\label{eqn:carrier_collision_time_gold}
\end{equation}
where $\tau_{e-e}^{-1}=AT_e^2$ and $\tau_{e-l}^{-1}=BT_l$~\cite{Vestentoft2006} are the scattering rates for electron-electron and electron-phonon interactions, respectively. In the future, the model could be improved 
by taking the electron heat capacity from detailed band structure calculations~\cite{Lin2008}.
The dielectric function can be written as,
\begin{eqnarray}
\epsilon_{\rm ex}({\bf r},t)&=&\epsilon_{\infty}+\epsilon_{\rm Drude}({\bf r},t),\\
\epsilon_{\rm Drude}({\bf r},t)&=&-{\frac{\left(\omega_p/\omega\right)^2}{1+i/\omega\tau_d({\bf r},t)}},
\label{eqn:epsilon_gold}
\end{eqnarray}
where $\varepsilon_\infty$ an is offset to the dielectric function that takes into account the effect of resonances at shorter wavelengths. Note in the case of gold $\omega_p=\sqrt{N e^2/m^*m_e\epsilon_0}$ is a constant but $\tau_d({\bf r},t)$ is locally and dynamically changing during the pulse, in contrast to the case of silicon.

\subsection{FDTD model}
\label{subsec:FDTD model}

To solve Eqs.~(\ref{eqn:maxwell_1}) and (\ref{eqn:maxwell_2}), we use a finite difference time 
domain (FDTD) method. As dictated by the Courant condition of the FDTD method (see for instance~\cite{Dennis2000}), 
the $E$ and $H$ fields are updated hundreds of times per optical cycle. Once every optical half-cycle,
we extract the electric-field amplitude $E_o({\bf r},t)$ from that optical half-cycle. 
Details of how we extract the amplitude can be found in Appendix~\ref{App:amp_extraction}. 
From the electric-field amplitude,
we determine the intensity $I({\bf r},t)$ using Eq.~(\ref{eqn:I_gen}). We use this intensity to
march Eq.~(\ref{eqn:carrier_gen}) and \ref{eqn:electron_temperature} (in the case of silicon) or 
Eq.~(\ref{eqn:electron_temperature_gold}) (in the case of gold) forward in time by a single step using an
implicit Euler method. As implicit Euler methods are unconditionally stable, we can choose the time step 
in the Euler method as half an optical cycle. 
The other quantities such as diffusivity, heat capacity, heat conductivity, etc. are subsequently calculated.
After this, we use the carrier density $N({\bf r},t)$ (in the case of silicon) or the carrier temperature $T_e({\bf r},t)$ (in the case of gold) to obtain a new dielectric function $\epsilon_{\rm ex}({\bf r},t)$. This updated dielectric function is used in the next optical half-cycle of the FDTD simulation. 

The self-reflectivity of a strongly focused laser pulse is in principle a three-dimensional problem.
However, three-dimensional finite difference time domain (FDTD) simulation are notoriously time and
memory consuming. We therefore instead run two-dimensional simulations for the TE and TM case and use
those to approximate the three-dimensional reflectivity. This requires an extra step that we discuss
at the end of this subsection. 
\begin{figure}[h]
\includegraphics[width=8.4cm]{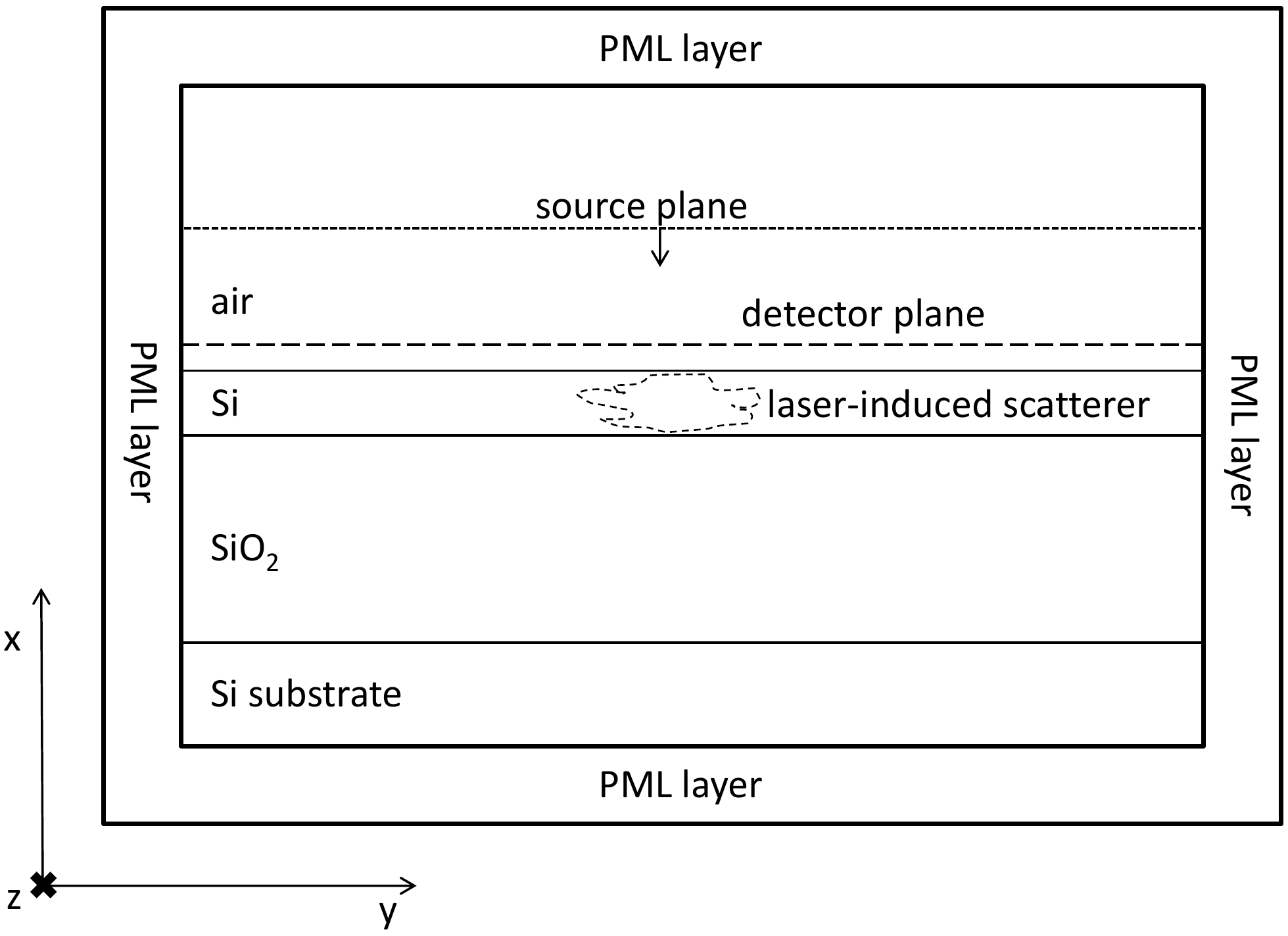}
\caption{Layout of the 2D-FDTD simulation box. The FDTD grid is excited by a soft source which is located $200~\rm nm$ above the silicon-air interface. Scattered $\bf E$ and $\bf H$ near-field values are recorded at the detector plane to extract the reflectivity.}
\label{fig:2D_FDTD_simulation_box}
\end{figure}
A schematic representation of our 2D-FDTD simulation box is shown in Fig.~\ref{fig:2D_FDTD_simulation_box}. The grid sizes for the simulations are chosen to be around $15~\rm nm$ such that the errors in the absolute reflectivity are smaller than 0.01 (See Appendix~\ref{App:FDTD_accuracy}) and the device layer thicknesses can be written as integers times the grid size. The width of the simulation box is $2~\rm \mu m$ which is two times the size of the focused laser spot ($\rm 1~\rm \mu m~@~1/e^2$ of intensity). The incident pulse duration used in the simulation is $126~\rm fs$, which is the value measured experimentally using a single-shot autocorrelator. The source plane is located at $200~\rm nm$ above the sample surface, while the near-field detector plane is located one cell above the sample surface. There are $30$ grid points in each of the four perfectly matched layers (PML) to ensure negligible reflections at the boundaries. We tested the accuracy of our method by comparing to several benchmarks, as discussed in 
Appendix~\ref{App:benchmark}. Due to the high carrier density in both gold and excited silicon, we need to
implement dispersion and loss in our FDTD method. Details of this implementation are given in 
Appendix~\ref{App:dispersive and lossy}.

To determine the field that will be scattered/reflected back by the sample, we run the simulation with 
and without the sample and take the difference in the electric field at the detector plane as the scattered 
near-field. The submicron-sized laser-induced plasma induces components with a spatial frequency, which is
too high to propagate into the far-field. We therefore filter the high spatial frequency components from the scattered near-fields, as described in Appendix~\ref{App:evanescent} in order to obtain the
scattered far-field. From the resulting fields, we calculate the reflected pulse fluence 
${\cal F}_{\rm refl}^{\rm TE,TM}(y)$ and the incident pulse fluence ${\cal F}_{\rm inc}^{\rm TE,TM}(y)$. 
Here, the superscripts TE and TM denote the results for the TE and the TM case. To obtain the total reflected/incident pulse energy, we add the TE and TM contributions. To approximate the three-dimensional 
results, we treat the $y$-coordinate in our simulation as the radial coordinate in a polar coordinate system
and integrate the reflected/incident fluence over the area of the incident focal spot
\begin{equation}
U_{\rm refl,inc}=\int_0^{r_{\rm max}} 2\pi r dr \left({\cal F}_{\rm refl,inc}^{\rm TE}(r)+{\cal F}_{\rm refl,inc}^{\rm TM}(r)\right),
\end{equation}
where $r_{\rm max}$ is chosen to be twice the waist of the focused laser spot.
Finally, we obtain the reflectivity
\begin{equation}
R=\frac{U_{\rm refl}}{U_{\rm inc}}.
\label{eqn:self_reflectivity_quasi_3D}
\end{equation}
This equation yields the value for $R$ to we will compare with experimental measurements in the next
section.

\section{Results and comparison with experiments}
\label{Sec:Exp_and_model}

In the case of silicon we have carried out simulations on thin-film, silicon-on-insulator (SOI) samples
as well as bulk silicon. Experimental results for these samples can be found in Ref.~\cite{Zhang2013}. For the simulations a number of input parameters need to be specified. The values we used are listed in Table~\ref{tab:simulation parameters}. In addition to these parameters, values for the Si and SiO$_2$ layer thicknesses
are required for the SOI samples. Table~\ref{tab:Fitted parameters} lists the parameters used in our simulations. As can be seen, we inserted values for the layer thicknesses slightly deviating from 
their measured values. These adjusted values were chosen in order to yield the correct self-reflectivity at 
vanishing fluence. It should be noted that the reflectivity in this regime depends only on the thicknesses of the layers and the refractive indices of the unperturbed media.

\begin{table}[h]
        \begin{tabular}{|l|l|l|}
        \hline
        Symbol&Description&Value\\\hline
        $\epsilon_{\rm Si}$&dielectric constant&$13.6+0.048i$~\cite{Choi2002}\\\hline
        $\beta~$&TPA coefficient&$\rm 1.85\times 10^{-9}~cm/W$~\cite{Bristow2007}\\\hline
        $n_2$&Kerr coefficient&$\rm 5\times 10^{-15}~cm^2/W$~\cite{Bristow2007}\\\hline
        $E_g$&Band gap&$\rm 1.12~eV$~\cite{Henry1987}\\\hline
        $\tau_d$&carrier collision time&$\rm 1.1~fs$~{\cite{Tinten2000}}\\\hline
        $\theta$&impact ionization coefficient&$\rm 21.2~cm^2/J$~{\cite{Pronko1998}}\\\hline
    \end{tabular}
    \caption{Material parameters used in the simulation as obtained from literature. The dielectric
constant refers to unexcited silicon at a wavelength of 800~nm.}
    \label{tab:simulation parameters}
\end{table}

\begin{figure}[h]
	\includegraphics[width=8.4cm]{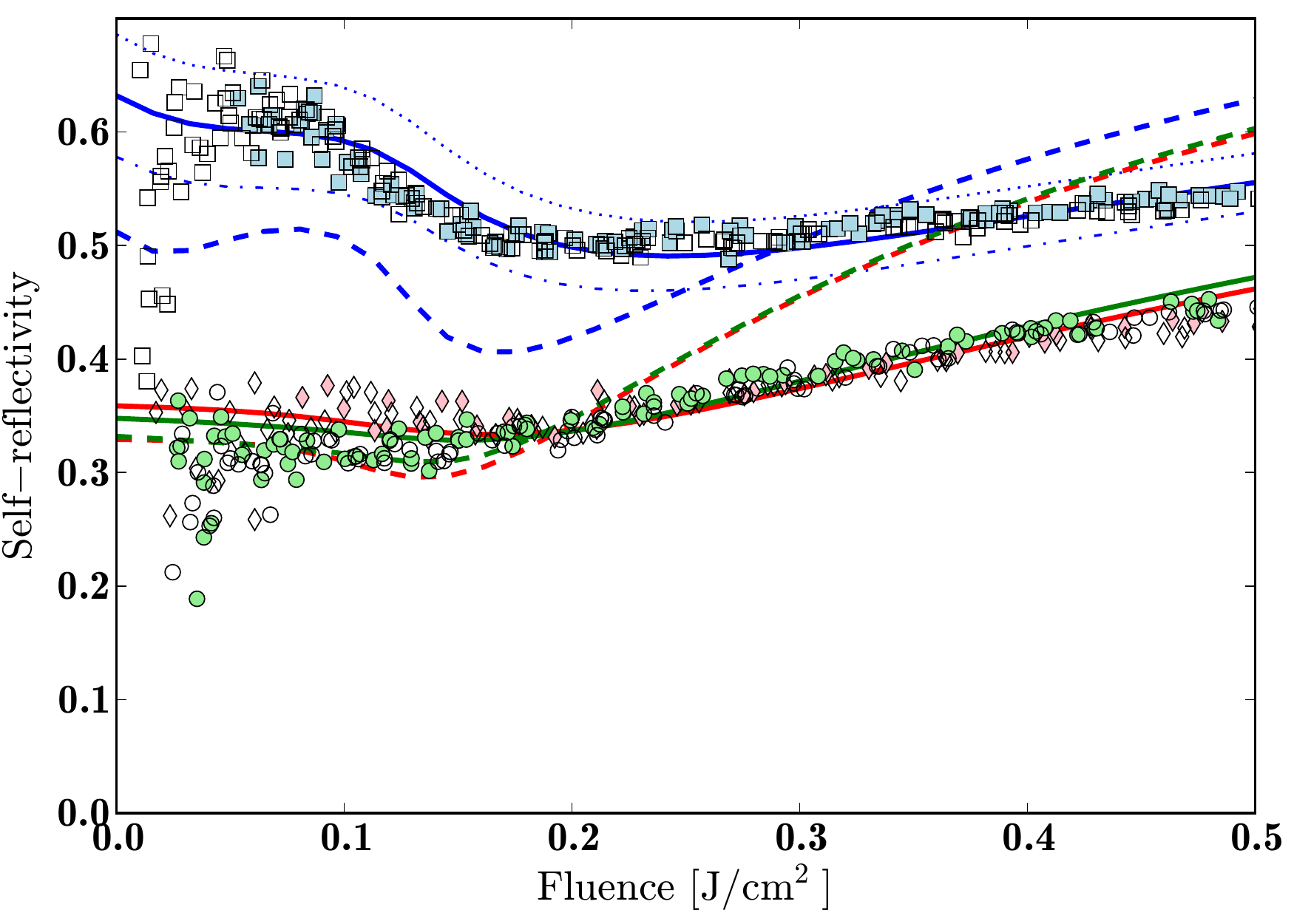}\\
	\caption{
	2D-FDTD calculations and experimental measurements of self-reflectivity for bulk silicon and 
	silicon-on-insulator (SOI) samples. Open and closed symbols indicate results of two independent 
	experimental runs for bulk silicon (circles), $\rm SOI_1$ (squares) and $\rm SOI_2$ (diamonds).
	The experimental errors are in all cases smaller than the symbols.
	The solid lines show the reflectivity calculated by 2D-FDTD simulations with the TM and TE modes 
	combined for bulk (green), $\rm SOI_1$ (blue) and $\rm SOI_2$ (red). The dashed lines show the
	results obtained from a 1D-FDTD simulation.
	The blue dash-dotted and dotted lines show the reflectivity calculated for $\rm SOI_1$ using either 
	the TE or the TM mode, respectively.}
\label{fig:self_reflectivity_and_model}
\end{figure}

Fig.~\ref{fig:self_reflectivity_and_model} shows the self-reflectivity calculated using the model and the experimental data of the bulk silicon and the SOI samples. As is shown for the $\rm SOI_1$ sample, the 
experimental data lie between the calculated self-reflectivity for the TM and the TE mode and agree well with the calculation considering both modes. For clarity, we do not show the TM and TE modes separately for the $\rm SOI_2$ and bulk samples. With slight changes in the values of $\theta$ and $\tau_d$ the agreement with the experimental data is even better~\cite{Zhang2013}. We see that the reflectivities for the bulk and the $\rm SOI_2$ samples are very similar, whereas the reflectivity of the $\rm SOI_1$ sample is very different from the other two samples. This is because the device layer of the $\rm SOI_1$ sample is thick enough to allow for constructive interference in the layer at 800~nm, impossible for the bulk sample
and the $100 \rm nm$ device layer of the $\rm SOI_2$ sample.
As the incident fluence increases, the reflectivity drops to a minimum and then increases again. This behavior suggests a typical free-carrier (Drude) response. As the carrier density increases, the real part of the refractive index first drops until it reaches the critical density (where the plasma frequency equals the incident light frequency), after which the real part of the refractive index increases again. The dashed lines in Fig.~\ref{fig:self_reflectivity_and_model} show the results of one-dimensional FDTD calculations using the same parameters as the two-dimensional calculations. The disagreement with the experimental data of the one-dimensional calculations directly shows that the wide range of incident angles must be taken into account when a high NA objective is used, as is the case in nano-ablation experiments.

\begin{table}[h]
        \begin{tabular}{|l|l|l|}
        \hline
        Parameter&Specified (Measured)&Adjusted\\\hline

        $d_1,{\rm SOI}_1$&$200~{\rm nm}$ ($201.3\pm 4.1~{\rm nm}$)&$200~{\rm nm}$\\\hline
        $d_2,{\rm SOI}_1$&$1000~{\rm nm}$&$970~{\rm nm}$\\\hline
        $d_1,{\rm SOI}_2$&$100~{\rm nm}$ ($111.5\pm 3.0~{\rm nm}$)&$100~{\rm nm}$\\\hline
        $d_2,{\rm SOI}_2$&$300~{\rm nm}$&$275~{\rm nm}$\\\hline
    \end{tabular}
        \caption{Sample parameters. The device layer thicknesses $d_1$ are measure using atomic force
microscopy. The specified values are also shown. For the buried oxide layer, only the specified values
are given. We use the specified parameters for the device layer thicknesses and adjust the buried oxide
layer to obtain the correct reflectivity in the low fluence limit.}
    \label{tab:Fitted parameters}
\end{table}

\begin{figure}[t]
   \centering
   \includegraphics[width=8.4cm]{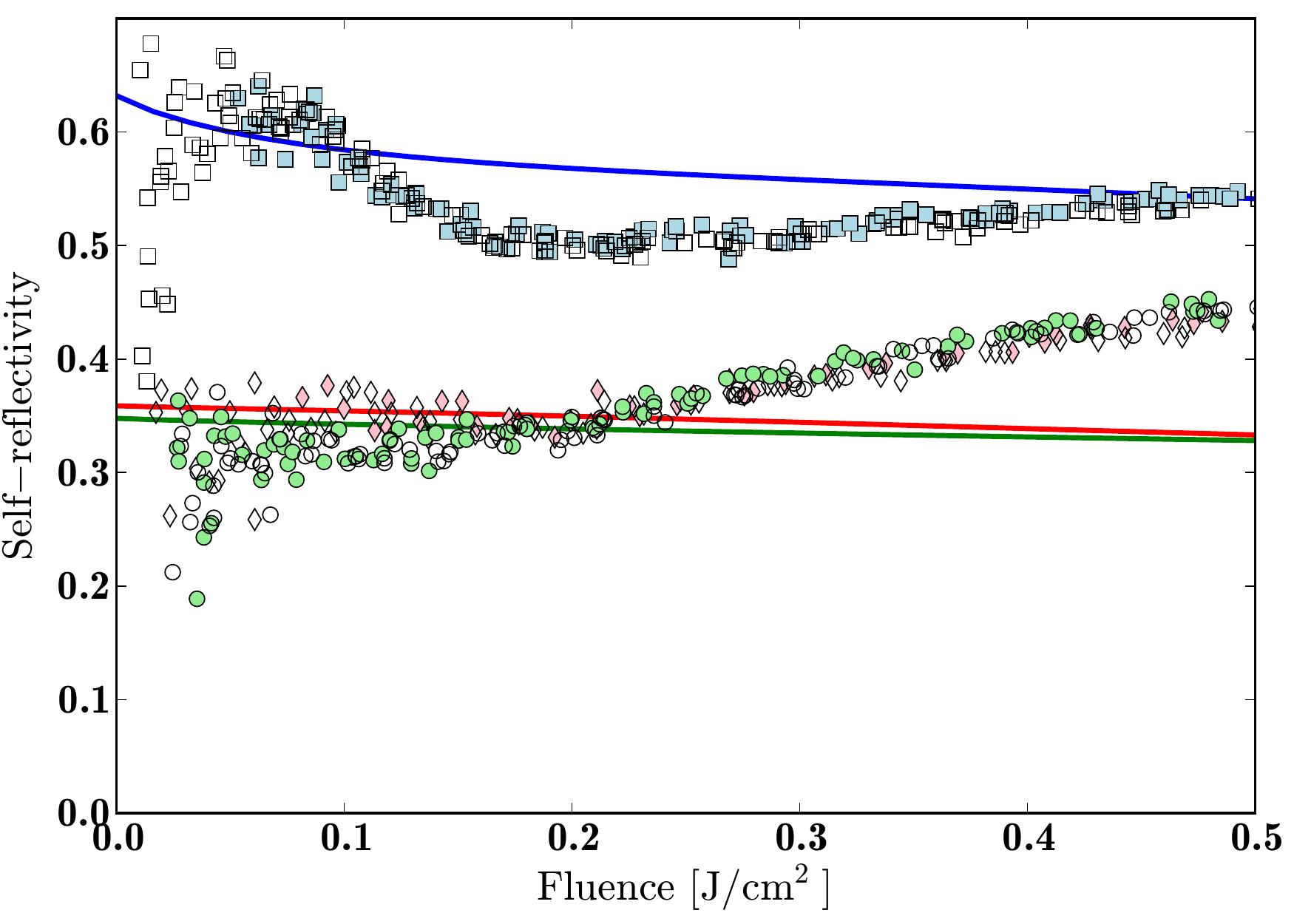}
   \caption{FDTD calculations and experimental measurements of self-reflectivity on the $\rm SOI_1$, $\rm SOI_2$ and bulk samples. For the calculations, an impact ionization coefficient $\rm\theta=0~cm^2/J$ is used. The
lines and symbols have the same meaning as in Fig.~\ref{fig:self_reflectivity_and_model}.}
\label{fig:self-reflectivity no_impact}
\end{figure}

\begin{figure}[htbp]
   \centering
   \includegraphics[width=8.4cm,height=5cm]{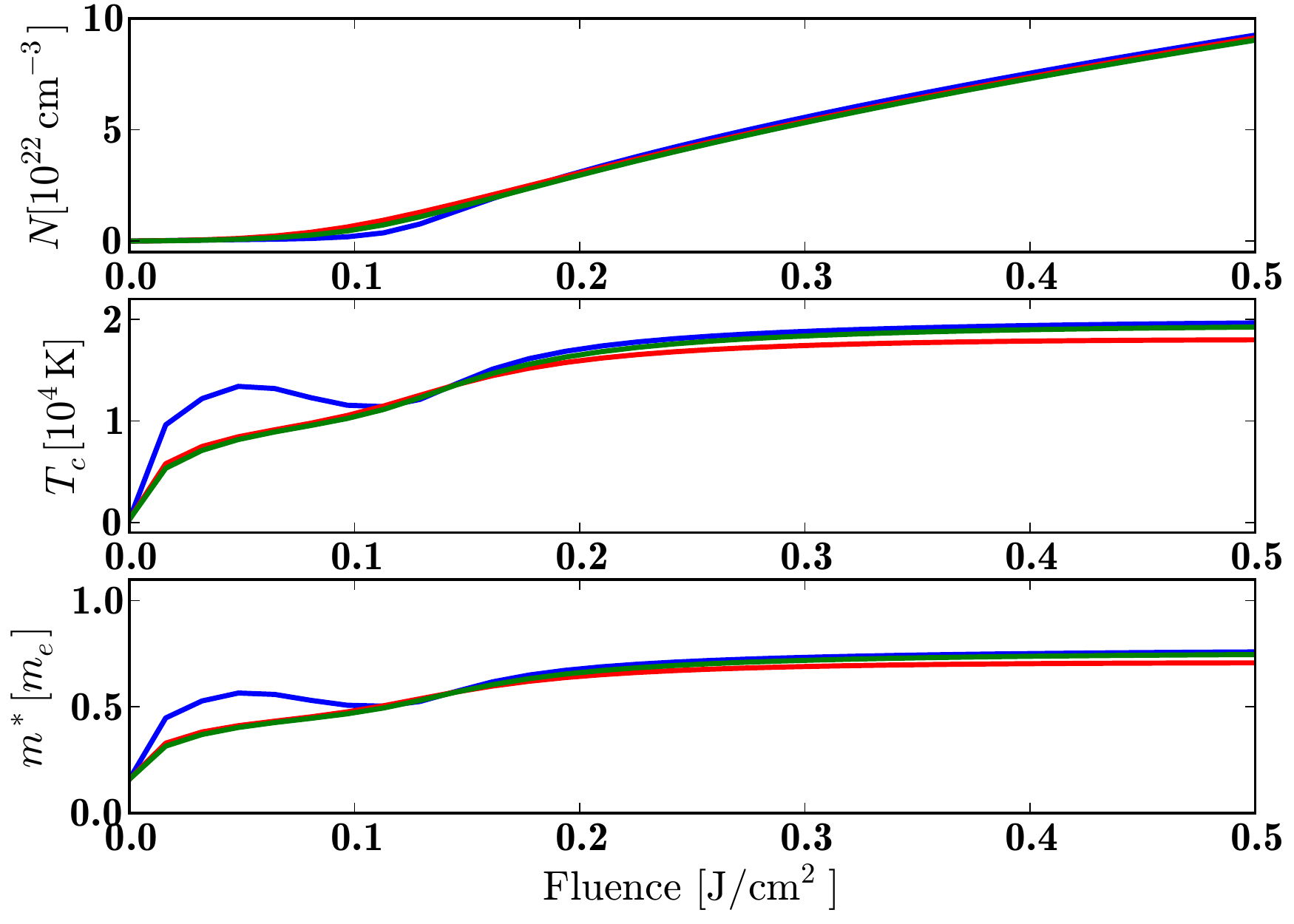}
   \caption{Calculated a) carrier density , b) carrier temperature and c) optical effective mass on the surface of the sample after the end of the pulse. In each plot the green line shows the data for bulk silicon sample, the blue line for the $\rm SOI_1$ sample and the red line for the $\rm SOI_2$ sample.}
\label{fig:N_surface}
\end{figure}

To elucidate the important role of impact ionization in the carrier creation process, we show in Fig.~\ref{fig:self-reflectivity no_impact} the calculation results with the impact ionization coefficient $\theta$ set to zero. The disagreement with the experimental data in Fig.~\ref{fig:self-reflectivity no_impact} and the excellent agreement with the experimental data in Fig.~\ref{fig:self_reflectivity_and_model} demonstrates directly that impact ionization plays a significant role in the development of the dense electron-hole plasma in silicon induced by a single femtosecond laser pulse.
It should be pointed out that in earlier work~\cite{Tinten2000} the role of impact ionization was ignored, which resulted in an underestimation of the carrier density and a fitted (static) optical effective mass of $m^*=0.18m_e$. However, as shown by Riffe's theoretical work, at a carrier temperature of $3000~\rm K$ the optical effective mass of silicon already exceeds $0.24m_e$~\cite{Riffe2002}. Considering that carrier temperature of above $10^4~\rm K$ are reached in the experiments~\cite{Tinten2000,Hulin1984}, the value $m^*=0.18m_e$ reported in Ref.~\cite{Tinten2000} is far too low. To see whether these conditions also occur in our model, we inspect the carrier density and the 
carrier temperature as calculated in our model. In Fig.~\ref{fig:N_surface}, we plot the results of those
quantities. Specifically, we plot the values obtained at the surface of the sample, directly after the pulse. In Fig.~\ref{fig:N_surface} (a) we find that the carrier density is 
clearly beyond $10^{22}~{\rm cm}^{-3}$ at the excitation level relevant for this work. 

In Fig.~\ref {fig:N_surface} (b), we see that the calculated carrier temperature is indeed larger than $10^4~
\rm K$ for all but the lowest fluences. Finally, in Fig.~\ref{fig:N_surface} (c), we plot the optical 
effective mass obtained from our model. It is also clearly beyond its unperturbed value ($0.15m_e$). This is caused by the high temperature reached in the laser-induced plasma. 
\begin{figure}[htbp]
   \centering
   \includegraphics[width=8.4cm]{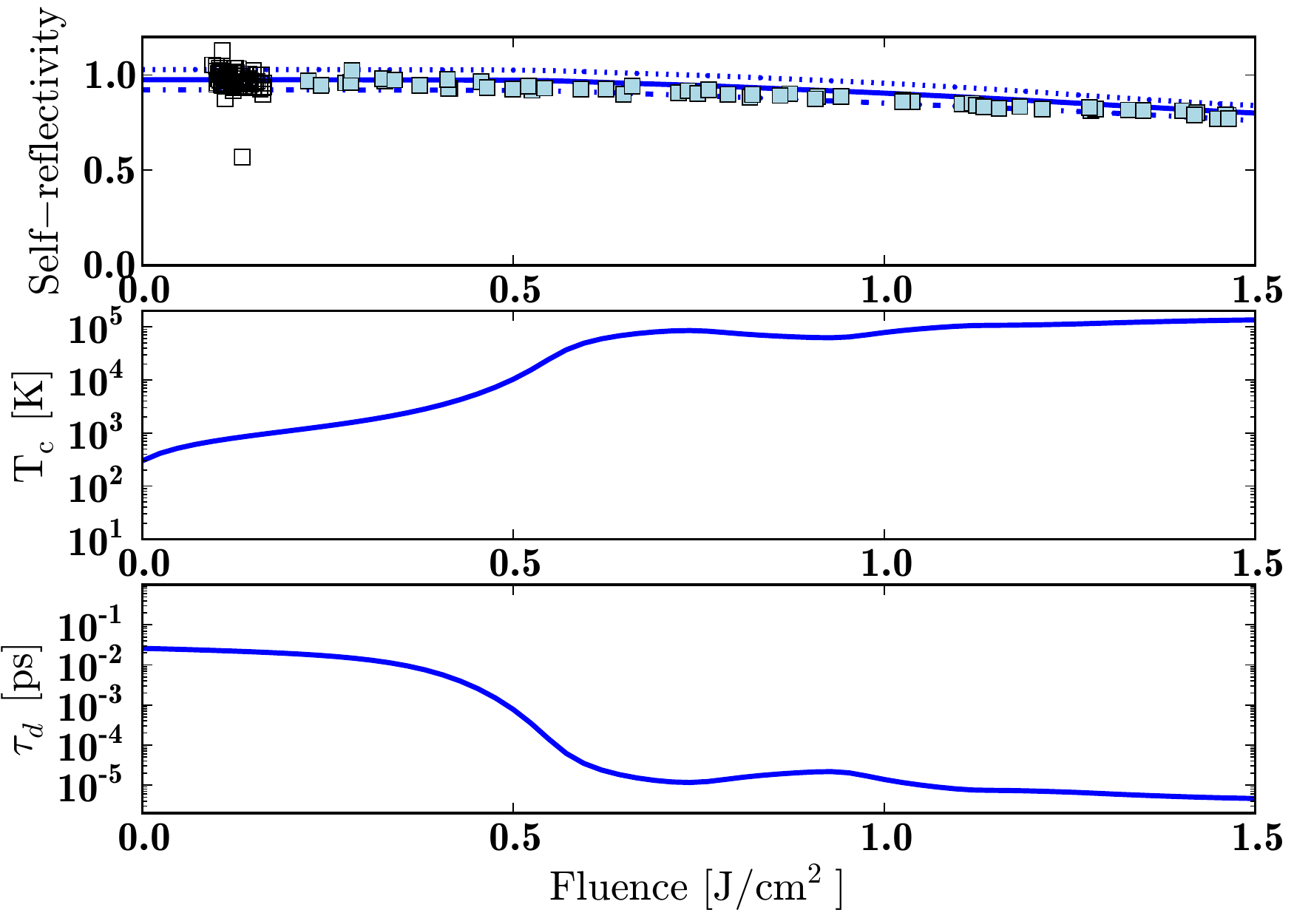}
   \caption{(a) FDTD calculation and experimental measurements of self-reflectivity on a $400~\rm nm$ gold film on glass. Open and closed squares are the experimental data from two independent runs. The solid line 
shows the calculated self-reflectivity taking both the TM and the TE mode into account. The dotted and dash-dotted lines show the self-reflectivity taking only the TM mode or only the TE mode into account, 
respectively. (b) Model calculated electron temperature on the surface of the sample. (c) The corresponding Drude damping time.}
\label{fig:self_reflectivity_and_model_gold}
\end{figure}

To test the versatility of our method, we also carried out experiments and simulations on a $400~\rm nm$ 
thick gold film grown on a glass substrate using vapor deposition. As mentioned in Sec.~\ref{subsec:LASER-MATTER INTERACTION FOR GOLD}, the case
for metals is simpler than that of semiconductors, as there is already such a high carrier concentration that
only the heating of those carriers has an influence on the optical properties of the samples. In Fig.~\ref{fig:self_reflectivity_and_model_gold} (a) we can clearly see this simplicity in the experimental results:
the self-reflectivity starts out high for low fluences and decreases slowly with increasing fluence. The
lines in the plot are the results from our model, where no free parameters where used. All parameters, as listed in Table~\ref{tab:simulation parameters_gold}, where taken from literature. We can see that the model 
excellently predicts the self-reflectivity from the unperturbed reflectivity without free parameters. 
The calculation also yields the electron temperature and the Drude damping time as shown in Fig.~\ref{fig:self_reflectivity_and_model_gold}(b) and (c), respectively. We find also in the case of gold,
temperatures beyond $10^4~\rm K$ for all but the lowest fluences. Surprisingly, we find in 
Fig.~\ref{fig:self_reflectivity_and_model_gold}(b) that the final carrier temperature $T_c$ at the surface
is not a monotonically rising function of incident fluence, but shows a slight decrease between fluence of $0.7$ and $1.0~\rm J/cm^2$. We attribute this reduction to the fact that the at temperatures of $T>10^4~\rm K$, the carrier damping time $\tau_d$ becomes shorter than the optical period. This means that above these temperatures, the imaginary part of the dielectric function and thus the absorption of the gold actually drop as a function of electron temperature.
    \begin{table}[htbp]
        \begin{tabular}{|l|l|}
        \hline
        Symbol&Value\\\hline
        $\epsilon_{\infty}$&$6$\\\hline
        $A$&$\rm 1.18\times 10^7~s^{-1}K^{-2}$~\cite{Vestentoft2006}\\\hline
        $B$&$\rm 1.25\times 10^{11}~s^{-1}K^{-2}$~\cite{Vestentoft2006}\\\hline
        $N$&$\rm 5.9\times 10^{22}~cm^{-3}$\\\hline
        $m^* m_e$&$\rm 1.1~m_e$~\cite{Vestentoft2006}\\\hline
    \end{tabular}
    \caption{Physical parameters used in the simulation for gold. The coefficients $A$ and $B$ are as defined in the text around Eq.~(\ref{eqn:carrier_collision_time_gold}), $N$ is the conduction electron density in gold, $m^*m_e$ is the optical 
effective mass and $\varepsilon_\infty$ is an offset to the Drude dielectric constant that takes resonances
at higher optical frequencies into account.}
    \label{tab:simulation parameters_gold}
    \end{table}

\section{Summary and conclusions}
\label{Sec:Summary}

We presented a model describing the propagation and absorption of a strongly focused femtosecond laser pulse used for single-shot laser ablation in semiconductor and metal samples, based on a two-dimensional 
FDTD method. 
The model is compared with self-reflectivity measurements of strongly focused femtosecond laser pulses used for single-shot femtosecond laser ablation on two SOI samples, bulk silicon and gold. We obtain excellent agreement between simulation and experiments, using the unperturbed reflectivity to adjust material and
sample specific constants; the self-reflectivity at high fluences follows without the use of adjustable 
parameters.
This confirms the accuracy and robustness of the model. The model clearly shows the dominant role of impact ionization for the carrier generation in silicon induced by a femtosecond laser pulse of $800~\rm nm$.
Furthermore, the model demonstrates a marked increase of the optical effective mass due to the elevation of the carrier temperature during the pulse. 

These results prove that FDTD simulations incorporating production and heating of free carriers and 
the free-carrier Drude response excellently describe the behavior of the self-reflectivity of strongly focused femtosecond laser beams under the ablation conditions. As the simulations accurately predict the 
self-reflectivity, it also gives a detailed understanding of the energy deposition of femtosecond laser pulses in metals and semiconductors. In conclusion, this extended FDTD method is an indispensible tool in the study of femtosecond laser nano-structuring of materials. 

\begin{appendix}
\section{Extracting the complex amplitude from FDTD simulation}
\label{App:amp_extraction}
The FDTD method calculates real-value,  time-varying electric and magnetic fields. However, some relevant 
physical quantities, such as the intensity Eq.(~\ref{eqn:I_gen}), are more conveniently expressed in 
the amplitude of the oscillation.
We extract the amplitudes of the fields from the simulation as follows. If we assume the amplitude is slowly 
varying with respect to the optical cycle, we can write the electric field on time $t$ as
\begin{equation}
E(t)=E_0 \cos(\omega_0 t+\phi)
\label{eqn:time_harmonic field},
\end{equation}
where $\omega_0$ is the frequency of the light, and $E_0$ and $\phi$ are the amplitude and phase, respectively. If we integrate $E^2(t)$ over half an optical cycle ${T/2}$, we find
\begin{eqnarray}
\int_0^{T/2}E^2(t)dt&=&\int_0^{T/2}E_0^2\cos^2(\omega_0 t+\phi)dt\nonumber\\
&=&\frac{1}{4} T E^2_0,
\label{eqn:time_harmonic field integration}
\end{eqnarray}
and thus
\begin{equation}
E_0=\sqrt{\frac{4\int_0^{T/2}E^2(t)dt}{T}}
\label{eqn:time_harmonic field integration2}.
\end{equation}
In the FDTD simulation, the integration is approximated as a summation
\begin{equation}
\int_0^{T/2}E^2(t)dt\cong\sum\limits_{n=n_1}^{n_1+m}E^2(n\Delta t)\cdot\Delta t,
\label{eqn:time_harmonic field integration_appro}.
\end{equation}
where $n_1$ is the starting time step of the summation and $m$ is the total time steps contained in half an optical cycle.

To extract the phase $\phi$ we consider the integral
\begin{equation}
\int_0^{T/2}E_0\cos(\omega_0 t+\phi)e^{-i\omega_0 t}dt=
\frac{E_0 \pi}{2\omega_0}\left(\cos\phi+i\sin\phi\right).
\label{eqn:integral_phase_extraction_1}
\end{equation}
Thus, the phase of the electric field is the phase of above integral.

\section{FDTD accuracy}
\label{App:FDTD_accuracy}

The FDTD method has intrinsic second-order accuracy because it uses central difference for both the time and space derivative. A second source of error of a FDTD code is the small residual reflection at the PML boundary. In this appendix, we analyze the numerical error due to the FDTD algorithm. Based on this result, we deduce the right grid size and PML thickness for the simulations performed in this paper. To analyze the error, we calculate the reflectivity of a plane wave under normal incidence using one-dimenional FDTD method and compare the result to the exact Fresnel result~\cite{BornWolf1999}
\begin{equation}
R=\left(\frac{n-1}{n+1}\right)^2,
\label{eqn:fresnel equation}
\end{equation}
which for silicon at $800~\rm nm$ incident wavelength yields a reflectivity of $0.3287$. The one-dimensional FDTD simulation space we use consists $50~\rm nm$ of vacuum and $1~\rm \mu m$ silicon, sandwiched between 
two PML layers. In order to extract the reflectivity, we carry out the simulations with and without the silicon layer. In the simulations we record the time-varying $E$ and $H$ field at the detector which is located one grid space above the silicon/vacuum interface. Thus the reflected time-varying Poynting vector is
\begin{equation}
S_{\rm ref}=(E-E_{\rm free})\times (H-H_{\rm free}),
\label{eqn:Poyning_Chapter2}
\end{equation}
where $E_{\rm free}$ and $H_{\rm free}$ are the fields from the simulation without the silicon layer. 
To obtain the reflectivity, we integrate the Poynting vectors $S_{\rm ref}$ over time
\begin{equation}
F=\int S_{\rm ref} dt,\,F_0=\int S_{\rm free} dt,
\label{eqn:fluence_Poyning_Chapter2}
\end{equation}
and write $R^\prime={F/F_0}$.
The absolute error ${\rm err}$ in the FDTD simulation is then defined as
\begin{equation}
{\rm err}=R^\prime-R.
\label{eqn:absolute error 1DFDTD}
\end{equation}
\begin{figure}[t]
\includegraphics[width=8.4cm]{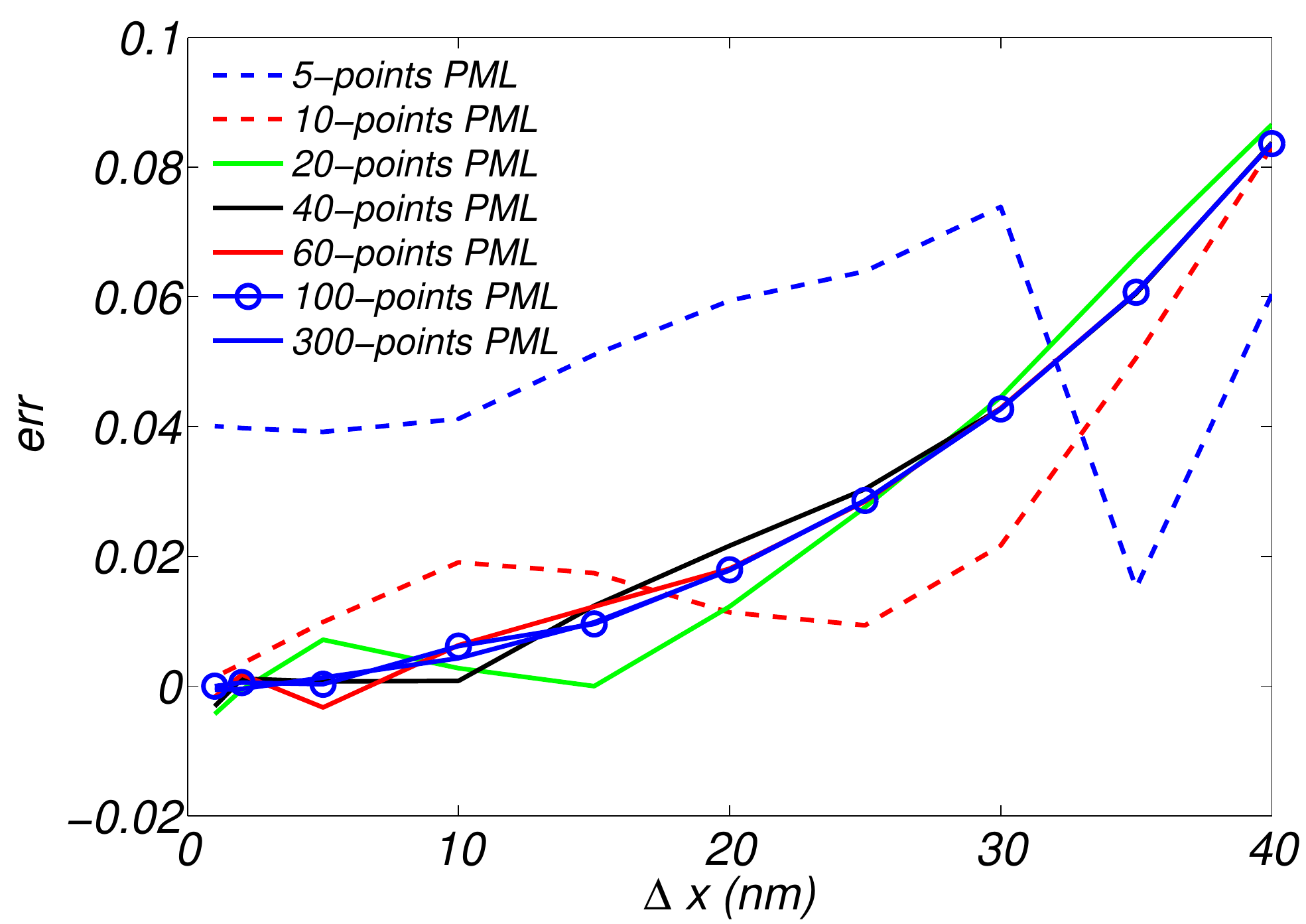}
\caption{The error of the FDTD method. Results with a range of PML thicknesses are presented.}
\label{fig:FDTD_error}
\end{figure}
We run the simulation with a range of different grid spacings $\Delta x$ and PML layer thicknesses $n_{\rm pml}$ measured in the number of cells in the PML layer. The pulse duration is set to $100~\rm fs$. The electric field of this pulse is used as a soft source~\cite{Dennis2000} to excite the grid. Fig.~\ref{fig:FDTD_error} shows the resulting absolute error in reflectivity. From the figure we can see that for a coarser grid, the error is very large even for the thickest PML layer while for the finest grid, decreasing the PML thickness gives rise to larger error. This suggests that for a coarser grid the error from central difference approximation dominates while for a fine grid, the error from the PML layer dominates. To make the FDTD simulation very accurate, one needs to choose a fine grid and at the same time choose a thick PML layer. From the figure, we conclude that in order to keep the error in the absolute reflectivity smaller than 0.01, one needs a grid spacing $\leq 15~\rm nm$ and a PML layer not less than $20$ points.

\section{Benchmark}
\label{App:benchmark}

\begin{table}[h]
        \begin{tabular}{|l|l|l|l|l|}
        \hline
        $\rm Structure$ &$R_{TM}$&$R_{TM}^{\prime}$&$R_{TE}$&$R_{TE}^{\prime}$\\\hline
        $\rm Bulk~Si$&0.322&0.338&0.345&0.356\\\hline
        $\rm SOI$&0.499&0.505&0.495&0.507\\\hline
        $\rm SOI^*$&0.220&0.237&0.236&0.252\\\hline
    \end{tabular}
    \caption{A comparison between the results of our 2D-FDTD code with the results from the commerical software FDTD Solutions. The results by FDTD Solutions are denoted by $R_{TM}^{\prime}$ and $R_{TE}^{\prime}$. The $*$ symbol represents the presence of a micron-sized scatterer embodied in the SOI wafer.}
    \label{tab:2DFDTD_benchmark}
\end{table}

We developed the 2D-FDTD code based on the method described in the next Appendix. As a test for the validity of the 2D-FDTD code, we calculate the reflectivity of both unperturbed SOI with a $230~\rm nm$ thick device layer and unperturbed bulk silicon with our FDTD code and compare the results with a commercial FDTD 
solver~\cite{FDTDSolutions}. The oxide layer of the SOI wafer is $1 \mu\rm m\rm$. 
We use a one-dimensional Gaussian profile of the electric field as a soft source to excite the simulation. 
The $\rm 1/e$ width of the Gaussian pulse is set to $100~\rm fs$. The grid size in the simulation is set as $13~\rm nm$. We extract the reflectivity from the field values calculated by the simulations. The results are summarized in Table~\ref{tab:2DFDTD_benchmark}. 
We further tested the code by calculating the reflectivity of a micron-sized scatterer (dispersive and lossy) embodied in the device layer of the SOI wafer. As can been seen, the results obtained by the FDTD Solutions and our FDTD code are very close to each other. The small differences are most likely due to differences in PML layer thicknesses and/or the grid spacing. 

\section{Dispersive and lossy media}
\label{App:dispersive and lossy}

In two dimensions, the Maxwell's equations reduce to independent equations for the TM and TE modes~\cite{BornWolf1999}. For the TM mode, these become~\cite{Dennis2000}
\begin{eqnarray}
\frac{\partial \tilde{D}_z}{\partial t}&=&\frac{1}{\sqrt{\epsilon_0\mu_0}}(\frac{\partial H_y}{\partial x}-\frac{\partial H_x}{\partial y}),
\label{eqn:FDTD_TM_Maxwell's equations 1_renormalized}\\
\tilde{D}_z(\omega)&=&\epsilon_r^*(\omega) \tilde{E}_z(\omega),
\label{eqn:FDTD_TM_Maxwell's equations 2_renormalized}\\
\frac{\partial H_x}{\partial t}&=&-\frac{1}{\sqrt{\epsilon_0\mu_0}}\frac{\partial \tilde{E}_z}{\partial y},
\label{eqn:FDTD_TM_Maxwell's equations 3_renormalized}\\
\frac{\partial H_y}{\partial t}&=&\frac{1}{\sqrt{\epsilon_0\mu_0}}\frac{\partial \tilde{E}_z}{\partial x}.
\label{eqn:FDTD_TM_Maxwell's equations 4_renormalized}
\end{eqnarray}
Whereas for the TE mode they become
\begin{eqnarray}
\frac{\partial \tilde{D}_x}{\partial t}&=&\frac{1}{\sqrt{\epsilon_0\mu_0}}\frac{\partial H_z}{\partial y},
\label{eqn:FDTD_TE_Maxwell's equations 1_renormalized}\\
\frac{\partial \tilde{D}_y}{\partial t}&=&-\frac{1}{\sqrt{\epsilon_0\mu_0}}\frac{\partial H_z}{\partial x},
\label{eqn:FDTD_TE_Maxwell's equations 2_renormalized}\\
\tilde{D}_x(\omega)&=&\epsilon_r^*(\omega) \tilde{E}_x(\omega),
\label{eqn:FDTD_TE_Maxwell's equations 3_renormalized}\\
\tilde{D}_y(\omega)&=&\epsilon_r^*(\omega) \tilde{E}_y(\omega),
\label{eqn:FDTD_TE_Maxwell's equations 4_renormalized}\\
\frac{\partial H_z}{\partial t}&=&-\frac{1}{\sqrt{\epsilon_0\mu_0}}(\frac{\partial \tilde{E}_y}{\partial x}-\frac{\partial \tilde{E}_x}{\partial y}).
\label{eqn:FDTD_TE_Maxwell's equations 5_renormalized}
\end{eqnarray}
In the above equations, we introduce scaler electric and displacement fields as
\begin{eqnarray}
\tilde{\mathbf{E}}&=&\sqrt{\frac{\epsilon_0}{\mu_0}}\mathbf{E},\\
\label{eqn:FDTD_normalising E}
\tilde{\mathbf{D}}&=&\frac{1}{\sqrt{\epsilon_0\mu_0}}\mathbf{D},
\label{eqn:FDTD_normalising D}
\end{eqnarray}
to make the electric and magnetic fields the same order of magnitude.

Note that Eqs.~(\ref{eqn:FDTD_TM_Maxwell's equations 2_renormalized}), (\ref{eqn:FDTD_TE_Maxwell's equations 3_renormalized}) and (\ref{eqn:FDTD_TE_Maxwell's equations 4_renormalized}) are expressed in the frequency domain whereas the FDTD method works in the time domain. To bring the frequency domain equations into the
time domain, we need to assume they are off a known analytical form. Here, we use a Drude model
\begin{eqnarray}
\epsilon_{\rm ex}&=&\epsilon_{\rm Si}+\chi_{\rm Drude}\nonumber\\
&=&\epsilon_{\rm Si}-\left(\frac{\omega_p}{\omega}\right)^2\frac{1}{1+i\frac{1}{\omega\tau}}.
\label{eqn:dielectric constant of excited silicon_CH5}
\end{eqnarray}
Using partial fraction expansion and switching the imaginary unit from $i$ to $j$ ($j=-i$, as is conventional in engineering), Eq.~(\ref{eqn:dielectric constant of excited silicon_CH5}) can be written as
\begin{eqnarray}
\epsilon_{\rm ex}&=&\epsilon_{\rm Si}+\frac{\omega_p^2\tau}{j\omega}-\frac{\omega_p^2\tau^2}{1+j\omega\tau}\nonumber\\
&=&\epsilon_{\rm Si}+\frac{\sigma_{\rm Drude}}{j\omega\epsilon_0}+\frac{\chi}{1+j\omega\tau}.
\label{eqn:dielectric constant of excited silicon_CH5_2}
\end{eqnarray}
Where we introduced the parameters
\begin{eqnarray}
\sigma_{Drude}&=&\epsilon_0\omega_p^2\tau,\\
\label{eqn:sigma_Drude}
\chi&=&-\omega_p^2\tau^2,
\label{eqn:chi_Drude}
\end{eqnarray}
where $\sigma_{\rm Drude}$ is the conductivity of the plasma. The $\chi$ term causes additional dispersion. This is referred to as the Debye formulation~\cite{Dennis2000} of the Drude model. Inserting this dielectric function, the electric displacement reads
\begin{eqnarray}
D(\omega)&=&\epsilon_{\rm ex}(\omega)E(\omega)\nonumber \\
&=&\epsilon_{\rm Si}E(\omega)+\frac{\sigma_{\rm Drude}}{j\omega\epsilon_0}E(\omega)+\frac{\chi}{1+j\omega\tau}E(\omega)\nonumber \\
&=&D^\prime(\omega)+S(\omega),
\label{eqn:D_debye_medium}
\end{eqnarray}
where the first two terms of the right-hand side of the equation are summarized as $D^\prime(\omega)$ and the last term is written as $S(\omega)$.

As the FDTD operates in the time domain, Eq.~(\ref{eqn:D_debye_medium}) must be transformed into the time domain. We first transform $D^\prime(\omega)$ into the time domain. Recall that $\frac{1}{j\omega}$ in the frequency domain corresponds to integration in the time domain, so $D^\prime(\omega)$ becomes
\begin{equation}
D^\prime(t)=\epsilon_{\rm Si}E(t)+\frac{\sigma_{\rm Drude}}{\epsilon_0}\int_0^t{E(t\prime)}dt\prime.
\label{eqn:sigma_drude_time_domain}
\end{equation}
This integral is approximated as a summation over the time steps $\Delta t$
\begin{equation}
D^{\prime n}=\epsilon_{Si}E^n+\frac{\sigma_{Drude}\Delta t}{\epsilon_0}\sum\limits_{i=0}^{n}E^i,
\label{eqn:sigma_drude_time_domain_sum}
\end{equation}
where $n$ indicates the time step at $t=n\Delta t$. In the FDTD algorithm, we use this equation
to determine the current $E^n$ from the current $D'^n$ and the previous values of $E$. We do this by 
first separating the $E^n$ term from the rest of the summation
\begin{equation}
D^{\prime n}=\epsilon_{Si}E^n+\frac{\sigma_{Drude}\Delta t}{\epsilon_0}E^n+\frac{\sigma_{Drude}\Delta t}{\epsilon_0}\sum\limits_{i=0}^{n-1}E^i,
\label{eqn:sigma_drude_time_domain_sum_2}
\end{equation}
and solving that equation for $E^n$
\begin{equation}
E^n=\frac{D^{\prime n}-\frac{\sigma_{Drude}\Delta t}{\epsilon_0}\sum\limits_{i=0}^{n-1}E^i}{\epsilon_{Si}+\frac{\sigma_{Drude}\Delta t}{\epsilon_0}}.
\label{eqn:sigma_drude_time_domain_sum_3}
\end{equation}

We now treat the $S(\omega)$ term in Eq.~(\ref{eqn:D_debye_medium}) in a similar manner. We convert the term
\begin{equation}
S(\omega)=\frac{\chi}{1+j\omega\tau}E(\omega),
\label{eqn:Chi_drude_frequency domain}
\end{equation}
to the time domain to find
\begin{equation}
S(t)=\frac{\chi}{\tau}\int_0^t e^{-\frac{t-t\prime}{\tau}}E(t^\prime)dt^\prime.
\label{eqn:Chi_drude_time domain}
\end{equation}
We approximate the integral as a summation
\begin{eqnarray}
S^n&=&\chi\frac{\Delta t}{\tau}\sum\limits_{i=0}^{n}e^{-\frac{\Delta t(n-i)}{\tau}}E^i\nonumber\\
&=&\chi\frac{\Delta t}{\tau}\left[E^n+\sum\limits_{i=0}^{n-1}e^{-\frac{\Delta t(n-i)}{\tau}}E^i\right].
\label{eqn:Chi_drude_time domain_sum_1}
\end{eqnarray}
We can add Eq.~(\ref{eqn:Chi_drude_time domain_sum_1}) to Eq.~(\ref{eqn:sigma_drude_time_domain_sum_2}) to find the electric displacement for all the three terms
\begin{eqnarray}
D^n&=&\epsilon_{\rm Si}E^n+\frac{\sigma_{\rm Drude}\Delta t}{\epsilon_0}E^n+\frac{\sigma_{\rm Drude}\Delta t}{\epsilon_0}\sum\limits_{i=0}^{n-1}E^i\nonumber\\
&+&\chi\frac{\Delta t}{\tau}[E^n+\sum\limits_{i=0}^{n-1}e^{-\frac{\Delta t(n-i)}{\tau}}E^i],
\label{eqn:sigma_and chi drude_time_domain_sum}
\end{eqnarray}
which when we solve for $E^n$ yields
\begin{equation}
E^n=\frac{D^n-\frac{\sigma_{\rm Drude}\Delta t}{\epsilon_0}\sum\limits_{i=0}^{n-1}E^i-\chi\frac{\Delta t}{\tau}\sum\limits_{i=0}^{n-1}e^{-\frac{\Delta t(n-i)}{\tau}}E^i}{\epsilon_{Si}+\frac{\sigma_{\rm Drude}\Delta t}{\epsilon_0}+\chi\frac{\Delta t}{\tau}}.
\label{eqn:sigma_and chi drude_time_domain_sum_2}
\end{equation}

As in the case of pure damping, we calculate $E^n$ (the current value of $E$) from the current value of $D$ and the summation of all previous values of $E$.

We can easily describe two-photon absorption (TPA) and the optical Kerr effect into Eq.~(\ref{eqn:sigma_and chi drude_time_domain_sum_2}) by introducing an extra conductivity term due to the TPA and an extra change in the real part of the dielectric constant due to the optical Kerr effect. Together with the one-photon absorption (OPA), Eq.~(\ref{eqn:sigma_and chi drude_time_domain_sum_2}) can finally be written as
\begin{equation}
E^n=\frac{D^n-\frac{\sigma\Delta t}{\epsilon_0}\sum\limits_{i=0}^{n-1}E^i-\chi\frac{\Delta t}{\tau}\sum\limits_{i=0}^{n-1}e^{-\frac{\Delta t(n-i)}{\tau}}E^i}{R\rm e\rm[\epsilon_{Si}]+\Delta \epsilon+\frac{\sigma\Delta t}{\epsilon_0}+\chi\frac{\Delta t}{\tau}}.
\label{eqn:sigma_and chi drude_time_domain_sum_3}
\end{equation}
Where $\sigma=\sigma_{\rm OPA}+\sigma_{\rm TPA}+\sigma_{\rm Drude}$ and $\Delta \epsilon$ is the change in the real part of the dielectric constant due to the optical Kerr effect. The term $\sigma_{\rm OPA}$ can be linked to the imaginary part of the dielectric constant of un-excited silicon. The terms $\Delta \epsilon$ and $\sigma_{\rm TPA}$ are linked to the real and imaginary part of $\chi^{(3)}$, respectively.

\section{Spatial filtering}
\label{App:evanescent}

In the simulation, interaction between the light and the localized laser-induced plasma gives rise to 
field components with high spatial-frequencies. When we decompose the wave vector into the transverse 
and the longitudinal component they are related by
\begin{equation}
k_x=\sqrt{k^2-k_y^2},
\label{eqn:wave_vector_k_kx_ky}
\end{equation}
where $k$ is the wavenumber in air and $k_y$ is the transverse wavenumber. We observe from the above equation that for $k_y^2>k^2$ the longitudinal wavenumber $k_x$ is purely imaginary, therefore the wave associated with it is evanescent and will not propagate into the far-field. Prior to calculating the reflectivity, these components should therefore be filtered out. 
The transverse wavenumber can be conveniently written as $k_y=k\sin\theta$, where $\theta$ is the angle of reflection. As the numerical aperture of the objective used in our experiments is $\rm NA=sin\theta=0.8$. This means that scattered waves associated with a transverse wavenumber $k_y>0.8k$ will not be collected by the objective.
This means that what is measured in Figs.~\ref{fig:self_reflectivity_and_model} and \ref{fig:self_reflectivity_and_model_gold} is in fact the scattered field associated with a transverse wavenumber $k_y<0.8 k$. So in the FDTD simulation, spatial-frequency components with $k_y>0.8 k$ should be filtered out. This is accomplished by the spectral decomposition fields~\cite{Novotny2006}
\begin{equation}
\widehat{A}(k_y)=\frac{1}{\sqrt{2\pi}}\int_{-\infty}^{\infty}A(y)e^{-ik_yy}dy,
\label{eqn:spectrum representation of the fields}
\end{equation}
and transforming this composition back using a truncated inverse Fourier transform
\begin{equation}
A^\prime(y)=\frac{1}{\sqrt{2\pi}}\int_{-0.8k}^{0.8k}\widehat A(k_y)e^{ik_yy}dk_y.
\label{eqn:spectrum representation of the fields_inverse}
\end{equation}

\end{appendix}

\begin{acknowledgments}
We thank Cees de Kok and Paul Jurrius for discussions and technical assistance. We thank Sandy Pratama
for proof-reading the manuscript.
HZ acknowledges the financial support from China Scholarship Council. DMK acknowledges support from NSF grant DMR 1206979.
\end{acknowledgments}

\end{document}